\documentclass[aps,prl,twocolumn]{revtex4}
\usepackage{graphicx}
\usepackage{epsfig}
\usepackage{subfigure}

\begin{document}

\title{Electronic Transport in Underdoped YBCO Nanowires: Possible Observation of Stripe Domains}

\author{J. A. Bonetti}
\thanks{Present address:  Department of Physics, California Institute of Technology, Pasadena, CA 91125}
\author{D. S. Caplan}
\author{D. J. {Van Harlingen}}
\author{M. B. Weissman}
\affiliation{Department of Physics and Materials Research
Laboratory, University of Illinois at Urbana-Champaign, Urbana, IL
61801}

\date{\today}

\begin{abstract}

\noindent We have measured the transport properties of a series of
underdoped YBa$_{2}$Cu$_{3}$O$_{7-\delta}$ nanowires fabricated
with widths of $100$-$250$ nm.  We observe large telegraph-like
fluctuations in the resistance between the pseudogap temperature
$T^*$ and the superconducting transition temperature $T_{c}$,
consistent with the formation and dynamics of a domain structure
such as that created by charge stripes.  We also find anomalous
hysteretic steps in the current-voltage characteristics well below
$T_{c}$.
\end{abstract}


\maketitle Despite years of intense study of the high-$T_{c}$
cuprates, many key questions remain unresolved, particularly
regarding the microscopic details of the pairing mechanism, the
nature of the pseudogap state, and the relationship between the
superconducting and pseudogap states.  One prominent category of
high-$T_{c}$ models involves the presence of a symmetry-breaking,
non-superconducting order parameter that competes or coexists with
superconductivity.  Examples of such ordering are various orbital
current models \cite{Chak,Lee,Varma} and charge stripe ordering
\cite{EK,Carlson-review,Kivelson-review,CastroNeto-review}. A
signature of underlying order would be the formation of domains
separating various phases or orientations of the order parameter.
To date, experimental evidence for such domains is suggestive but
inconclusive.  Neutron scattering data \cite{Mook1,Mook2,Ichi} and
transport measurements \cite{Ando} have been interpreted as
support for charge ordering and stripe formation. STM spectroscopy
imaging has revealed electronic phase separation into mesoscopic
scale superconducting and pseudogap domains \cite{Lang}, and
stripe features that have been attributed to charge stripes
\cite{Kapitulnik-stripes}, quasiparticle interference fringes
\cite{Davis-stripes}, and incommensurate spin-density waves
\cite{Yazdani-stripes}.

In this Letter, we report transport measurements on nanoscale YBCO
samples that show large amplitude switching noise in the
resistance at temperatures above the superconducting transition
temperature $T_{c}$ but below the pseudogap temperature $T^*$, and
hysteretic voltage steps in the current-voltage characteristics
well below $T_{c}$.  We argue that the resistance fluctuations may
be a signature of the formation and dynamics of a
symmetry-breaking domain structure arising from charge ordering
into stripes.  We consider whether the voltage steps could arise
from a similar mechanism or are caused by heating or phase slips
in the nanowire.

\begin{figure}[b]
\includegraphics[width=6.5cm,bb=50 90 600 720]{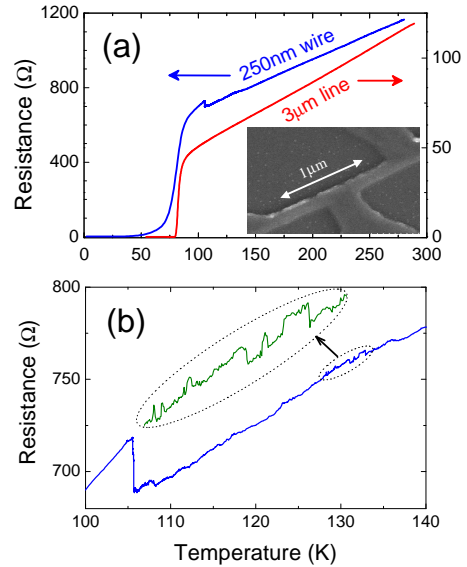}
 \caption{(a) Resistance vs. temperature of a $500$ nm segment of a
 $250$nm wide nanowire compared to that of a $3 ~\mu$m line.
 (b) Expanded section showing telegraph-like switching
 fluctuations at temperatures below approximately $150 K$.}
 \label{YBCO-nw-1}
\end{figure}

The key to the experiments is the fabrication of nanowires without
significant degradation of their superconducting properties.  We
start with YBCO thin films (thickness $50$-$100$ nm), grown by
pulsed laser deposition on LaAlO$_3$ (LAO) substrates, with
$T_{c}$'s from $\sim 92$ K (optimal doping) to $\sim 60$ K on the
underdoped side, as determined by a two-coil inductive
measurement. We then dc sputter a carbon layer ($\sim 200$ nm) and
a thin gold layer ($\sim {20}$ nm) to serve as an etching template
and to protect the film during processing. Patterning is done by
electron beam lithography and ion beam etching on a liquid
nitrogen cooled stage to avoid film degradation.  Finally, the
protective layers are removed and standard photolithography is
used to attach leads to the nanowire.

Four-point resistance measurements on short ($\sim 500$ nm)
segments of the nanowires revealed surprising results. The overall
shape of the resistance vs. temperature curves is as expected for
underdoped films: as the temperature is lowered, the typical
linear decrease in resistivity becomes sublinear below a
temperature $T^*$ normally associated with the pseudogap
temperature, followed by the transition into the superconducting
state. This is shown in Figure \ref{YBCO-nw-1}(a) for both a $3~
\mu$m line and a $250$ nm wide nanowire.  The transition width is
typically $3$-$5$ K with an onset at most a few degrees below the
transition temperature of the starting uniform film. However,
between a temperature $\sim T^*$ and $T_c$, the nanowire exhibited
large switching fluctuations of order $0.5$-$5 \%$ in the
resistance, as expanded in Figure \ref{YBCO-nw-1}(b). The largest
discrete resistance jump is nearly a $4\%$ change, and is preceded
by smaller ($\sim 1\%$) fluctuations. Also of interest are the
change in slope of R(T) and the absence of large fluctuations
immediately after the large $4\%$ switch.  Such switches are not
seen in the wider line.

\begin{figure}
 \includegraphics[width=5.5cm,bb=20 10 285 240]{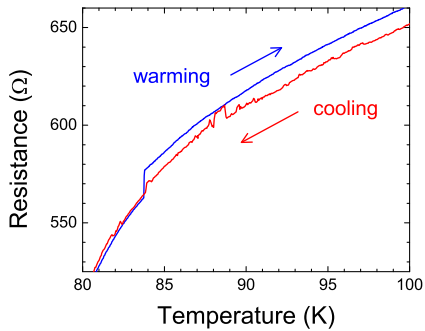}
 \caption{Successive cooling and warming resistance vs. temperature
 for a $250$ nm wide nanowire.  Between $87$-$89$ K, the cooling
 sweep exhibits multiple-level switching, with the high resistance state matching that
 established in the warming run.}
 \label{YBCO-nw-2}
\end{figure}

Data from another sample shown in Figure \ref{YBCO-nw-2}
demonstrates the behavior of the fluctuations upon successive
cooling and warming data runs.  Some features in the two runs are
correlated, such as the jumps in the resistance at $84$ K, while
others appear only when heating or cooling. The switching
fluctuations seen in the cooling curve appear to be telegraph-like
two-state or multiple-level switching with the highest resistance
state achieved corresponding to that measured during the the
warming curve. Simultaneous measurements on adjacent segments
suggest that most of the fluctuations are localized over regions
$\leq 100$ nm, but in a few cases correlations over extended
length scales $\geq 1~\mu$m were also observed.

We have observed such switching in more than five different
nanowire samples.  All of the samples studied are slightly
underdoped with $T_{c}$'s of $70$-$80$ K.  Thus, the pseudogap
crossover temperature $T^*$ should be roughly $50$-$100$ K above
$T_c$. This is indeed the case in those samples in which we can
identify $T^*$ as the temperature at which $R(T)$ deviates from a
linear variation. We observe the large resistance fluctuations in
approximately the same temperature range above $T_c$ and never
above $200$ K. Thus, although $T^*$ and the onset temperature of
the switching noise are difficult to determine clearly in some
nanowires, it is our impression that the large resistance
fluctuations occur only in the temperature range between $T_c$ and
$T^*$, suggesting that they may be a feature of the pseudogap
regime. We have not made a systematic study vs. sample width, but
larger samples of width $3$-$5 ~\mu$m do not display large
resistance fluctuations.

\begin{figure}[b]
 \centering
 \includegraphics [width=9cm, bb=40 320 800 660] {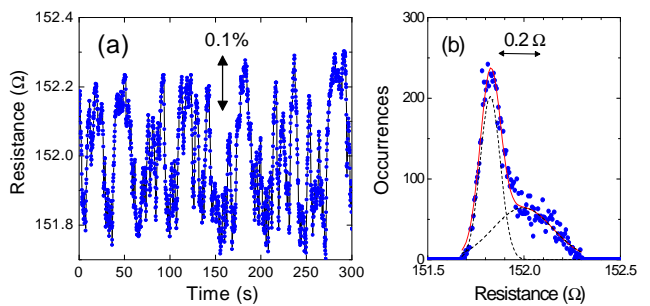}
 \caption{(a) Resistance vs time at $100$ K exhibits large $0.25\%$,
telegraph-like fluctuations. (b) Histogram of resistance values
demonstrating bimodal behavior.}
 \label{YBCO-nw-3}
\end{figure}

If the temperature is held fixed and the resistance is monitored,
it is sometimes possible to observe fluctuations in real time. An
example is shown in Figure \ref{YBCO-nw-3}(a).  The fluctuations
with amplitude $\approx 0.25\%$ have rather slow dynamics with
switching times on the order of one second. The sample appears to
prefer two states but, unlike a true two-level system,
intermediate metastable states are also present. A histogram of
the resistance states demonstrating a bimodal distribution is
shown in Figure \ref{YBCO-nw-3}(b).  Time traces taken at higher
temperatures well above $T^*$ did not exhibit bimodal switching.

Switching fluctuations in the resistance of conducting wires with
the large magnitude and the extended spatial range that we observe
are rare and difficult to explain. One possible mechanism is the
motion of extended defects such as dislocations which could affect
conduction channels. However, in this case it is difficult to
explain why the large fluctuations are seen only in the
temperature span $T_{c}<T<T^{*}$.  Other explanations would seem
to require the existence of a fluctuating domain structure with a
conductivity asymmetry or anisotropy.

One mechanism that incorporates adequately all of the observed
phenomena is the formation of charge stripes characterized by a
local anisotropy in the conductance.  Since the nanowires studied
are aligned with the $a$ or $b$ crystalline axes, such stripes
would naturally lead to a structure of domains aligned either
parallel or perpendicular to the measuring current flow. The two
orientations would have different critical currents, normal state
resistances, and current-voltage characteristics, with parallel
sections exhibiting a 1D wire behavior and perpendicular sections
described by tunneling between charge stripes.  If these domains
were to move, reorient, and/or change size, as would be expected
at higher temperatures, resistance fluctuations would result. If
we assume a typical domain size given by the stripe correlation
length $\Lambda \approx 35$-$40$ nm \cite{Mook2}, then the
nanowire segments we have measured would contain roughly a $5$ x
$20$ array of domains. Thus, a large resistivity change in one
such domain would result in a nanowire resistance change of order
$1$\%, in agreement with the observed values.

The two most challenging aspects of the data to explain are the
delocalized nature of the resistance changes and the slow
switching kinetics observed.  The stripe model provides a natural
potential explanation for the extended spatial correlations since
each Cu-O$_2$ plane of the layered YBCO structure may exhibit
stripe alignment independent of the others.  Alternatively, it has
been suggested that cuprate nanowires may exhibit edge states that
would show up as extended domains in the transport properties
\cite{Castro_Neto}. The edges may also affect the stripe dynamics
by forcing a preferred stripe domain alignment and pinning domain
wall motion, hence slowing the kinetics down to the timescales we
observe.  We find no evidence that the stripe domains are aligned
by current flow since the current-voltage (IV) characteristics of
the wire in the normal state are found to be strictly linear up to
currents of $100 ~\mu$A.

\begin{figure}
 \centering
 \includegraphics[width=6.5cm,bb=20 20 280 220]{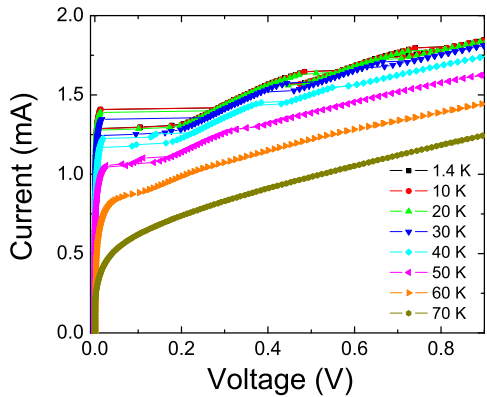}
 \caption{Current-voltage characteristics of a nanowire at different
 temperatures showing a phase-diffusion regime at low voltage and multiple
 hysteretic steps at higher current levels.}
 \label{YBCO-nw-4}
\end{figure}

Intriguing features are found in the IV curves at low
temperatures, as shown in Figure \ref{YBCO-nw-4} for a nanowire
segment of length $500$ nm and width $200$ nm measured at
different temperatures. As the current is increased we find the
onset of a finite phase diffusion voltage at a critical current
$I_{c}$, followed by two or more hysteretic steps.  The linear
portions after each step have a large excess current and do not
extrapolate to the origin, indicating that they are not simply
localized ohmic sections in the wire but instead likely involve
dynamical phase evolution. Simultaneous IV's on different but
overlapping segments indicate that the first large step is always
extended over several microns as in the normal state resistance
measurements, but other steps may be either extended or localized
in a single segment of the sample. Although this behavior could
suggest more than one mechanism for dissipation in the wire, the
different features may also arise from different vertical (c-axis)
layers of the sample.

We have found it difficult to explain the voltage steps by
standard mechanisms for the onset of resistance in superconducting
wires. We have considered explanations based on weak links between
grains, vortex flow, phase dynamics (phase slip centers), heating,
and defect dynamics and find that they cannot readily account for
the multiple voltage steps and their spatial correlations.

An obvious candidate for steps in the current-voltage
characteristics is weak links at localized spots in the wire. The
primary argument against these is that the low temperature
critical current density measured in our wires is very high
($>10^7$ A/cm$^2$), comparable to that of uniform thin films.
Films grown by pulsed laser deposition have a high density of twin
and low-angle grain boundaries, but these exhibit strong coupling
rather than Josephson behavior, giving large $J_{c}$ values of the
order we observe \cite{Mannhart}.  An explanation of our IV
results based on vortex entry and flux flow can also be ruled out.
The flux flow resistivity in a magnetic field $B$ scales as
$\rho_{f} = \rho_n (B/H_{c2})$, where $\rho_{n}$ is the normal
state resistivity and $H_{c2}$ is the upper critical field. For
the cuprates $H_{c2}$ is very large ($>100$ T), thus for the small
fields generated by the currents in our nanowires, $\rho_{f}$ will
be too low by several orders of magnitude to produce the linear
segments of the nanowire IV's observed between voltage steps.
However, flux flow does likely account for the onset of voltage in
the phase diffusion regime just above $I_c$.

Phase slip centers, the one-dimensional equivalent of dynamical
vortex flow, only occur in samples with widths $\lesssim \xi$, the
superconducting coherence length.  For YBCO, $\xi \sim 2$ nm so
our samples are substantially larger, with widths $\approx 50-100
~\xi$.  Thus, phase slip centers are an unlikely explanation.
However, a similar phenomena termed phase-slip lines may occur
even in wider lines \cite{Sivakov}. These may be thought of
resulting from the flow of kinematic vortices \cite{Andronov},
with properties intermediate between Abrikosov and Josephson
vortices.  These could be responsible for localized steps in the
IV characteristics but cannot easily explain their extended
nature.

A serious concern in measurements on current-carrying
nanostructures is local heating, which can cause localized regions
of the sample to exceed the critical temperature or critical
current \cite{Lavrov}. It is plausible that the steps in the
current-voltage characteristics could arise from this mechanism.
However, heating may not be as serious a problem in our cuprate
samples as in other nanostructures. Because of the high $T_c$, a
substantial local temperature rise is required to induce normal
transitions at the lowest measurement temperatures. At the same
time, cuprate films are extremely well thermally-anchored to the
substrate because LAO has a reasonably good thermal conductivity
\cite{Morelli} and the thermal boundary resistance between the
sample and substrate is very low due to their structural
similarity \cite{Nahum}.  To estimate the local temperature rise
from ohmic heating, we have modelled the heat flow from a nanowire
into the substrate, assuming that the sample attains a
steady-state temperature at each current-voltage point. For a
nanowire of width $w$ and length $L>>w$, the local temperature at
the center of the sample for uniform power dissipation $P=IV$ is
approximately given by
\begin{eqnarray}
T(P)&=& \left\{ T_{base}^2 + \frac{P}{\pi  \alpha L} ~ln
\left[\frac{(L^2+w^2)^{1/2}+L}{(L^2+w^2)^{1/2}-L}\right]
\right\}^{1/2}
\end{eqnarray}
where $T_{base}$ is the sample temperature with no current.  This
expression is valid in the temperature regime for which the
thermal conductivity of the substrate has the linear form
$\kappa=\alpha T$; in other regimes, the nanowire temperature must
be computed numerically.

\begin{figure}[b]
 \centering
 \includegraphics[width=6.5cm,bb=20 20 280 220]{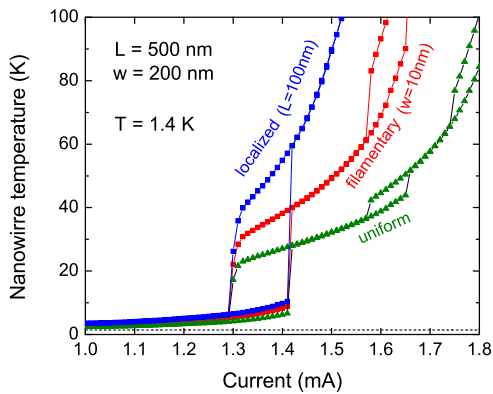}
 \caption{Sample temperature as a function of current for the nanowire IV
 shown in Fig. \ref{YBCO-nw-4} at $T=1.4$ K, assuming uniform, filamentary,
 and localized dissipation.}
 \label{YBCO-nw-5}
\end{figure}

As an example, we consider the nanowire IV at $T_{base}=1.4$ K
shown in Fig. \ref{YBCO-nw-4}, for which $L = 500$ nm and $w=200$
nm. The thermal conductivity of LAO is linear in temperature for
$T<25$ K with $\alpha=2.08$ W/m-K$^2$, peaks at $T \sim 30$ K, and
then drops roughly as $T^{-1/2}$.  We plot the calculated nanowire
temperature as a function of the current $I$ in Fig.
\ref{YBCO-nw-5}. The nanowire temperature heats substantially at
the highest currents, especially after the temperature becomes
high enough that the thermal conductivity drops, but the
temperature at the first observed voltage jump is predicted to be
less than $5$ K. Inhomogeneities in the current-flow and or
conductance of the wire will enhance the local heating.  We
simulate these possibilities by considering a narrower wire with
$w=10$ nm, corresponding to a filamentary conduction path, and a
shorter resistive section with $L=100$ nm, corresponding to a
localized resistive section.  The temperature rise in these cases
is larger but never exceeds $12$ K at the first jump, well below
the sample $T_c$.  Thus, although the heating is substantial in
our nanowires, it does not obviously explain the observed voltage
steps.  We note that the calculated temperature rise for the
normal state resistance measurements is negligible at the power
levels used ($\sim 0.1$-$10 ~\mu$W).

It is conceivable that the step structure could also be a
manifestation of the stripe domain structure suggested by the
resistance data.  If there are a series of domains aligned
parallel and perpendicular to the current direction, different
domains could switch into the normal state at different currents,
leading to multiple steps in the nanowire current-voltage
characteristics. Once a normal section is created, the heating
described above could drive large sections of the nanowire into
the normal state.  A detailed model has not been developed for
this scenario.

In conclusion, we have fabricated and measured the transport
properties of underdoped YBCO nanowires, giving a direct picture
of what occurs on mesoscopic length scales.  We observe discrete
switching noise in the resistance of the nanowires in the range
$T_{c}<T<T^{*}$.  This may be a signature of the formation of a
domain structure of charge stripes, which introduces a local
conductance anisotropy and allows domain dynamics.

We thank Eduardo Fradkin and Nigel Goldenfeld for insightful
discussions. This work supported by the U. S. Department of
Energy, Division of Materials Sciences, under Award No.
DEFG02-91ER45439, through the Frederick Seitz Materials Research
Laboratory at the University of Illinois at Urbana-Champaign. One
of us (JAB) thanks the Center for Nanoscale Science and Technology
at the University of Illinois for fellowship support.

\bibliography{stripes}

\end{document}